\documentclass[aps,groupedaddress,a4paper,twocolumn,prb,showpacs]{revtex4-1}
\usepackage{graphicx}
\usepackage{amsmath}
\usepackage{amssymb}
\usepackage{verbatim}
\usepackage{color}
\usepackage{dsfont}
\usepackage{bm}
\usepackage{hyperref}
\usepackage{mathtools}
\usepackage[normalem]{ulem}

\definecolor{violet}{rgb}{0.7,0,0.5}
\definecolor{newgreen}{rgb}{0,0.6,0.0}
\definecolor{grey}{rgb}{0.4,0.4,0.4}

\usepackage{amsthm}

 \theoremstyle{definition}

 \theoremstyle{remark}

\begin{document}

\title{Methodology for bus layout for topological quantum error correcting codes}
\author{Martin Wosnitzka}
\author{Fabio L. Pedrocchi}
\author{David P. DiVincenzo}
\address{JARA Institute for Quantum Information, RWTH Aachen University, D-52056 Aachen, Germany}
\begin{abstract}
Most quantum computing architectures can be realized as two-dimensional lattices of qubits that interact with each other. We take transmon qubits and transmission line resonators as promising candidates for qubits and couplers; we use them as basic building elements of a quantum code. We then propose a simple framework to determine the optimal experimental layout to realize quantum codes. We show that this engineering optimization problem can be reduced to the solution of standard binary linear programs. While solving such programs is a NP-hard problem, we propose a way to find scalable optimal architectures that require solving the linear program for a restricted number of qubits and couplers. We apply our methods to two celebrated quantum codes, namely the surface code and the Fibonacci code.
\end{abstract}


\maketitle

\section{Introduction}
Since the theoretical demonstration of fault-tolerant quantum information processing, a holy grail of modern physics has been to realize fault-tolerant quantum computing architectures in the lab. While this still remains a very challenging task, many experimental advances have been achieved. Arguably, one can hope to see the first small-size implementations in a near future.

Among the most promising quantum computing platforms, one finds so called topological quantum codes.\cite{TerhalReview} The main idea is to encode quantum information (in the form of logical qubits) using a large number of physical qubits. The additional degrees of freedom introduced in the Hilbert space then allow the extraction of some information about the errors induced by the environment (the error syndrome) and to correct them without collapsing the stored logical qubit. Furthermore, topological codes are, by definition, immune to local and static perturbations.\cite{TerhalReview}

Most of the topological quantum codes are realizable as a lattice of qubits (some of them might require qudits instead) that are coupled to each other. Depending on the specifics of the quantum code, one qubit might be coupled to several other qubits in its neighborood. In this work, we present a general framework to determine the optimal architecture to couple the qubits of a quantum code. Here we assume that couplers can be introduced between qubits and we identify the coupling architecture that minimizes the total length of the couplers, rendering the physical implementation more practical. Our analysis is valid for any quantum code and we show that this set of optimization problems are identical to well-known binary linear programs. 

We apply our formalism to two celebrated quantum codes, namely the surface code \cite{BravyiKitaev, FreedmanMeyer} and the Fibonacci Levin-Wen code. \cite{KKR,BDV} The former one is a planar version of Kitaev's toric code \cite{KitaevToric} that is among the most promising quantum computing platforms because of its simplicity and its surprisingly high error threshold of about 1\%. The former code is more involved but supports Fibonacci anyons that are universal for topological quantum computation; in other terms every quantum gate can be approximated to any accuracy by braiding Fibonacci anyons.

We think that our work on the surface code is particularly timely since the first set of experiments to build small fragments surface code (with 9 data qubits) have now started.\cite {Kelly2015} It is thus interesting to understand what architecture is optimal and could be realized in the lab.  Finally we compare our results for the surface code with previously suggested architectures.\cite{DiVincenzo,Ghosh} 

The paper is organized as follows. In Sec.~\ref{sec:Model} we present the physical model under consideration for a generic quantum code as well as the formalization of the optimization problem. In particular, we show that the optimal architecture is found by solving binary linear programs. In Sec.~\ref{sec:FLW} we apply the formalism developed in Sec.~\ref{sec:Model} in order to find an optimal architecture for the Fibonacci code. In particular, we present a methodology to find scalable architectures by solving tractable binary linear programs. Section~\ref{sec:SC} finally contains our results for the surface code.

\section{Connecting qubits optimally}\label{sec:1}
In this work, we consider Transmon Qubits (TQs) and Transmission Line Resonators (TLRs)  as the prototypical examples of physical qubits and moderate distance couplers.\cite{Blais,DiCarlo}
However, it is worth pointing out that our approach does not depend on the technological details of the implementation but can be applied to any kinds of qubits and couplers.\cite{Childress,Burkard,Trif,Trifunovic,Shulman,Trifunovic2,Hassler} 

\subsection{Model}\label{sec:Model}

\begin{figure}[h!]
	\centering
		\includegraphics[width=0.4\textwidth]{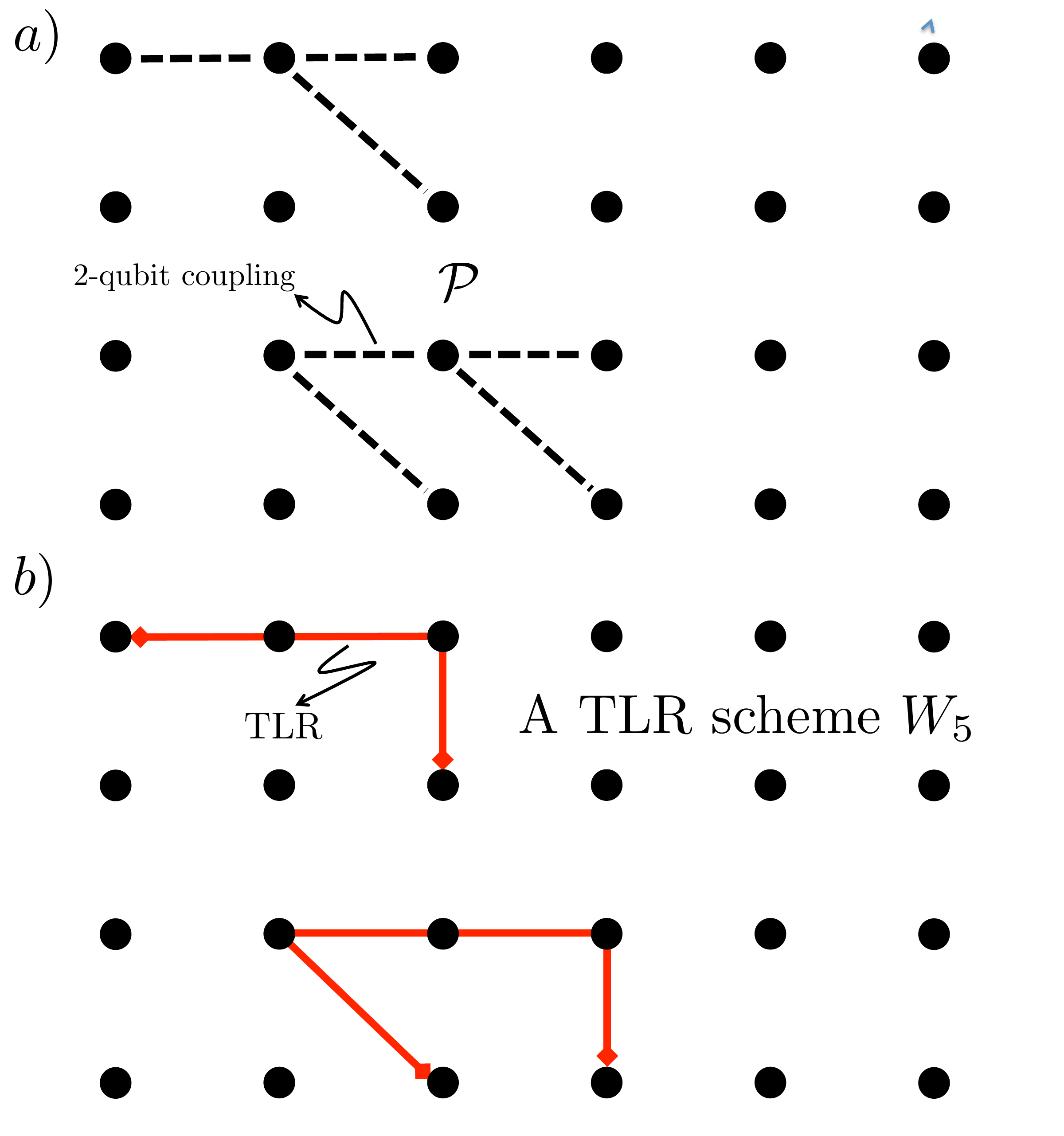}
	\caption{Two-dimensional lattice of qubits (black dots). The dashed lines between qubits represent two-qubit couplings. The solid lines going through the qubits are TLRs. The squares at the end of the solid lines specify the starting and the ending points of the corresponding TLRs. a) Set $\mathcal{P}$ of two-qubit couplings required in a given quantum circuit. b) TLR scheme that realizes every two-qubit coupling of a). The two TLRs each contain more than two TQs.}
	\label{fig:Protocole}
\end{figure}

Consider a set of $N$ TQs $q_{1},\ldots,q_{N}$ that lie on a two-dimensional plane at positions ${\bf x}_{1},\ldots, {\bf x}_{N}$. Depending on the specific quantum codes that one wants to realize, see Secs.~\ref{sec:Applications} for examples, several TQs must interact with each other and thus be coupled through TLRs. As any quantum circuit can be reduced to a succession of single- and two-qubit operations,\cite{NielsenChuang} the most straightforward approach is to introduce TLRs containing each exactly two TQs;  in this way TLRs realize the set $\mathcal{P}$ of all two-qubit couplings necessary to implement a given quantum circuit, see Fig.~\ref{fig:Protocole}. 

However, using a new TLR for each pair of qubits that should be coupled to participate in two-qubit gate operations might not be the most optimal approach to this engineering problem. In fact TLRs are able to couple to more than two TQs and we assume that $m$ individual TQs can reside inside the resonant cavity provided by the TLR. Each of the TQs can be controlled separately and coupled to any of the other $m-1$ TQs through the TLR. Following recent experimental progress,\cite{Chow} we find that $m\leqslant 5$ is a realistic upper bound. Also, it seems natural to restrict the number $p$ of TLRs that are connected to a single TQ; here we choose $p=5$. \cite{Riste} 

We call an unordered sequence of sites $i_{k}\in\{1,\ldots,N\}$ a string $\mathcal{S}=\{i_{1},i_{2},\ldots, i_{m}\}$. The length $\vert \mathcal{S}\vert$ of a string is defined by the number of sites it contains. To each string $\mathcal{S}$, we associate a number $\kappa_{\mathcal{S}}=0,1$; if $\kappa_{\mathcal{S}}=1$, then a TLR is present and hosts the $m$ TQs  $q_{i_{1}},\ldots,q_{i_{m}}$, otherwise no single TLR hosts all those specific $m$ qubits. We denote by $\mathfrak{S}_{m}$ the set of all strings $\mathcal{S}$ with $\vert\mathcal{S}\vert\leqslant m$. We call the vector
\begin{eqnarray}
W_{m}&=&(\kappa_{\{1,2\}},\ldots,\kappa_{\{N-1,N\}};\nonumber\\&&\hspace{1cm}\kappa_{\{1,2,3\}}, \ldots;\kappa_{\{1,2,\ldots,m\}},\ldots)^{T}
\end{eqnarray}
a TLR scheme.  We say that a TLR corresponding to a string $\mathcal{S}$ is \emph{included} in a TLR scheme $W_{m}$ if $\kappa_{\mathcal{S}}$ is one of the elements of the vector $W_{m}$. We say that a TLR scheme $W_{m} \emph{contains}$ a TLR associated with string $\mathcal{S}$ if it is included and $\kappa_{\mathcal{S}}=1$. 

In order to formalize the concept of \emph{optimal} TLR scheme, we introduce a cost $\mathcal{C}_{\mathcal{S}}\in\mathbb{R}$  associated with each string $\mathcal{S}\in\mathfrak{S}_{m}$. The cost vector of the scheme $W_{m}$ is then
\begin{eqnarray}
\mathcal{C}(W_{m})&=&(\mathcal{C}_{\{1,2\}},\ldots,\mathcal{C}_{\{N-1,N\}},\nonumber\\
&&\mathcal{C}_{\{1,2,3\}},\ldots,\mathcal{C}_{\{1,2,\ldots,m\}},\ldots,)^{T}.
\end{eqnarray}
Using this notation, the total cost of a given TLR scheme $W_{m}$ is $\mathcal{C}(W_{m})^{T}\cdot W_{m}$.   
The goal of this work is to determine one TLR scheme $W_{m}$ that minimizes the cost and realizes the set $\mathcal{P}$ of two-qubit couplings. In Secs.~\ref{sec:Applications} we present concrete examples of cost functions for the surface code and for the Levin-Wen model.

The problem of finding the TLR scheme $W_{m}$ that has minimal cost is solved by using standard binary linear optimization methods. The problem is formalized as follows: 
\textbf{Given a set of two-qubit connections $\mathcal{P}$, and given two integers $m$ and $p$, find the TLR scheme $W_{m}$ that minimizes the cost $\mathcal{C}(W_{m})^{T}\cdot W_{m}$ such that
\begin{enumerate}
\item For all $i_{j}\in\{1,2,\ldots,N\}$, \begin{equation}\sum_{\mathcal{S}\in\mathfrak{S}_{m}\,\vert\, i_{j}\in\mathcal{S}}\kappa_{\mathcal{S}}\leqslant p\,.\end{equation}
\item $W_{m}$ realizes every two-qubit coupling of $\mathcal{P}$.
\end{enumerate}}
It is worth pointing out again that the maximal number $m$ of qubits per TLR, as well as the maximal number $p$ of TLR per qubit, is fixed.

It is now clear why we call this a \emph{binary} linear program; every component of the vector $W_{m}$ is either 0 or 1. Solving such a binary linear program is generally very difficult and is in fact an NP-hard problem. However, specific instances of such problems can be tractable, and we give explicit examples below.  As a side remark, note that when all the numbers in the program are allowed to be real, then the situation is dramatically simplified and the optimization problem can be solved in polynomial time.

In this work, we use the free software lpsolve, available at \url{http://lpsolve.sourceforge.net/5.5/}, to find the optimal solution to the binary linear program defined above.  In order to simplify the program, we leave out all the \emph{superfluous} TLRs. We call a TLR superfluous if it can be replaced by two (or more) TLRs that host no common qubits such that the same set of required two-qubit couplings is realized; one can thus always replace a superfluous TLR by two TLRs that will have a lower overall cost. 

As mentioned in the Introduction, we aim to find the optimal architectures for two important quantum error correcting codes, namely the surface code and the Levin-Wen model. We find interesting that such quantum technological problems can be turned into standard optimization problems.

\section{Application to Quantum error correcting codes}\label{sec:Applications}

\subsection{Fibonacci Levin-Wen model}\label{sec:FLW}
Levin-Wen models are a class of spin systems defined on trivalent lattices whose excitations realize any consistent (abelian or non-abelian) anyonic theory.\cite{LevinWen} Here we focus on a particular Levin-Wen model, namely the Fibonacci Levin-Wen model.\cite{KKR,BDV} Its name takes its origin in the nature of the excitations above the ground states; indeed they are Fibonacci anyons with topological charge $\tau$ and fusion rules
\begin{equation}
\tau\times\tau=1+\tau\,.
\end{equation}
Here $1$ represents the vacuum topological charge.

\begin{figure}[h!]
	\centering
		\includegraphics[width=0.4\textwidth]{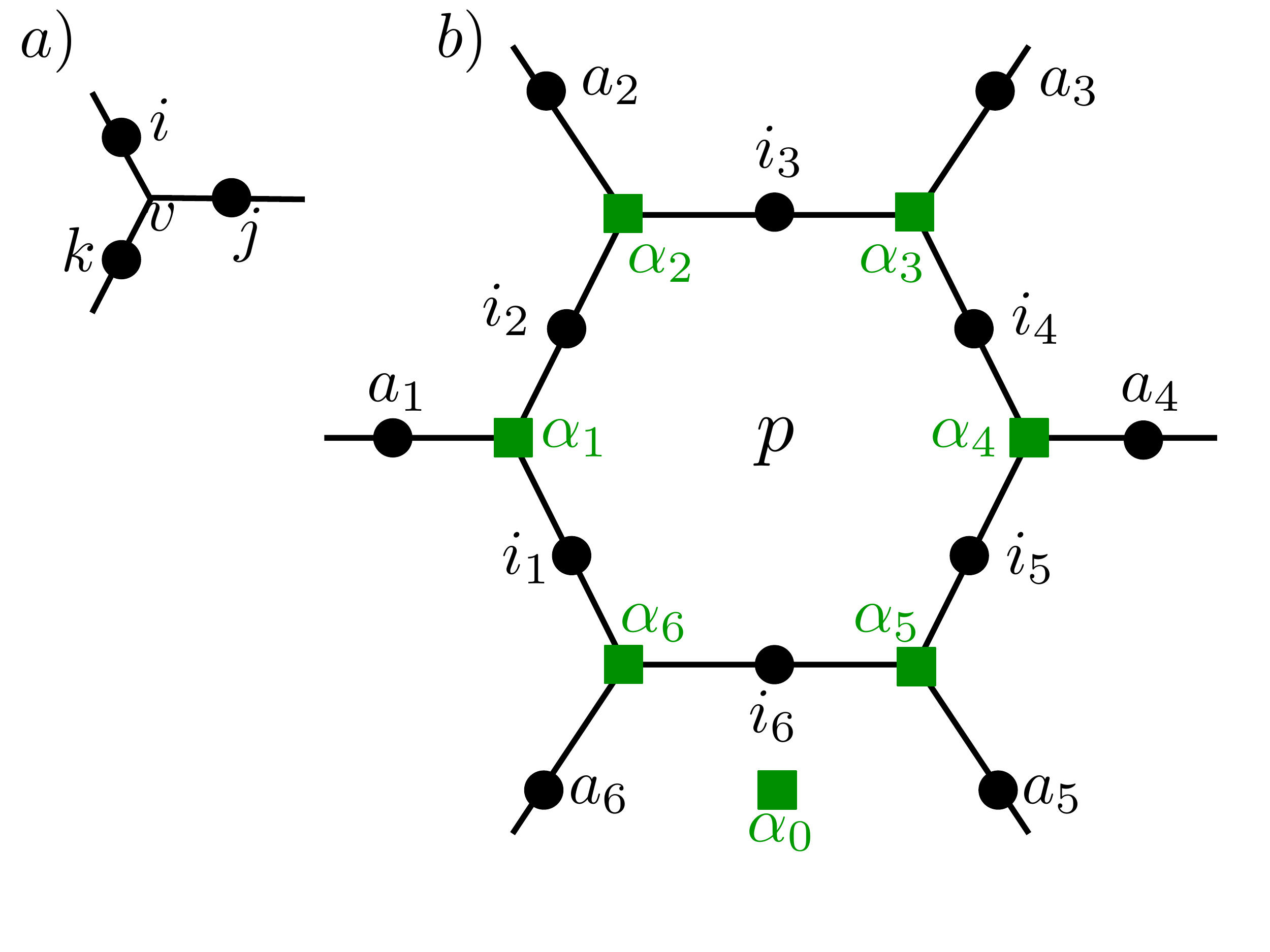}
	\caption{The Fibonacci Levin-Wen model is defined on a trivalent lattice. Each edge hosts a spin-$1/2$ particle (a so-called data qubit) depicted here by a black dot. $a)$ Vertex $v$ where three edges of the lattice meet. The state of the three qubits at the vertex $v$ is $\vert ijk\rangle$. $b)$ Twelve data qubits (black dots) needed to define the plaquette operator $B_{p}$ on the trivalent lattice. In order to perform non-demolition measurements of vertex and plaquette operators, one introduces ancillary qubits (green squares). Here $\alpha_{0}$ is used to measure $B_{p}$, while the remaining ancillary qubits $\alpha_{1-6}$ are used to measure the six vertex operators. This number of additional qubits is appropriate for the plaquette reduction method of Ref.~\onlinecite{BDV}.}
	\label{fig:LW}
\end{figure}

Considering a trivalent lattice with each edge carrying a spin-$1/2$ particle, we define the Fibonacci Levin-Wen Hamiltonian,\cite{KKR,BDV}
\begin{equation}\label{eq:LWH}
H=-\sum_{v}Q_{v}-\sum_{p}B_{p}\,,
\end{equation}
where $Q_{v}$ and $B_{p}$ are operators that are respectively associated with vertex $v$ and plaquette $p$ of the lattice, see Fig.~\ref{fig:LW}. 

The vertex operator $Q_{v}$ acts on the three qubits residing on the edges that meet at vertex $v$. If the states of the three qubits on theses edges are $\vert i\rangle$, $\vert j\rangle$, and $\vert k\rangle$, then we have
\begin{equation}
Q_{v}\vert ijk\rangle=\delta_{ijk}\vert ijk\rangle\,,
\end{equation}  
with
\begin{equation}
\delta_{ijk}=\left\{\begin{array}{cc}
1 & \text{if $ijk=000,011,101,110,111$,}\\
0 & \text{otherwise}\,.
\end{array}\right.
\end{equation}
The plaquette operators are more complicated and involve 12-qubit interactions. Consider the twelve qubits $a_{1-6}$ and $i_{1-6}$ around a given plaquette $p$, see Fig.~\ref{fig:LW}b). The plaquette operators are then defined through
\begin{equation}
B_{p}=\frac{1}{1+\phi^{2}}\left(B_{p}^{1}+\phi B_{p}^{\tau}\right)\,,
\end{equation}
with $\phi=\frac{1+\sqrt{5}}{2}$ the golden ratio and
\begin{eqnarray}
B_{p}^{s}\vert a_{1},\ldots,a_{6},i_{1},\ldots,i_{6}\rangle&=&\sum_{i^{\prime}_{1},\ldots,i^{\prime}_{6}}B_{p,i_{1},\ldots,i_{6}}^{s,i^{\prime}_{1},\ldots,i^{\prime}_{6}}(a_{1},\ldots, a_{6})\nonumber\\
&&\times\vert a_{1},\ldots,a_{6},i^{\prime}_{1},\ldots,i^{\prime}_{6}\rangle\,,
\end{eqnarray}
where $s=1,\tau$ and
\begin{equation}
B_{p,i_{1},\ldots,i_{6}}^{s,i^{\prime}_{1},\ldots,i^{\prime}_{6}}(a_{1},\ldots,a_{6})=F^{a_{1}i_{6}i_{1}}_{s i^{\prime}_{1}i^{\prime}_{6}}F^{a_{2}i_{1}i_{2}}_{si^{\prime}_{2}i^{\prime}_{1}}\cdots F^{a_{5}i_{4}i_{5}}_{s i^{\prime}_{5}i^{\prime}_{4}}F^{a_{6}i_{5}i_{6}}_{si^{\prime}_{6}i^{\prime}_{5}}\,.
\end{equation}
For the Fibonacci theory we have \cite{Bonderson}
\begin{eqnarray}
F^{\tau\tau\tau}_{\tau}&=&\begin{pmatrix}
				\phi^{-1} & \phi^{-1/2}\\
				\phi^{-1/2} & -\phi^{-1}
				\end{pmatrix}\,,
\end{eqnarray}
and all other $F$'s are trivial.
One can then show that the Levin-Wen plaquette and star operators satisfy $\left[B_{p},Q_{v}\right]=[B_{p},B_{p^{\prime}}]=[Q_{v},Q_{v^{\prime}}]=0$, for all $v,v^{\prime},p,p^{\prime}$.\cite{KKR,BDV}

We define the Fibonacci code \cite{KKR} $\mathcal{F}$ (an example of a stabilizer code) as the ground-state subspace of Hamiltonian (\ref{eq:LWH}), namely
\begin{equation}
\mathcal{F}=\{\vert\psi\rangle\, \vert\,  Q_{v}\vert\psi\rangle=B_{p}\vert\psi\rangle=\vert\psi\rangle, \forall\, p,v\}\,.
\end{equation}

On a surface with nontrivial topology this ground-state subspace of Hamiltonian (\ref{eq:LWH}) is degenerate and one uses this set of states to encode logical qubits. 
A nontrivial operation (a logical error) applied to the logical qubit is implemented by creating pairs of $\tau$-excitations, braiding them, and annihilating them. The logical operation does not depend on the details of the braiding process, but only on its topology; this is in fact the main idea of topological quantum computation.\cite{ReviewTQC} Importantly, Fibonacci anyons are universal for quantum computation and any quantum gate can thus be performed in a topologically protected fashion.

\begin{figure}[h!]
	\centering
		\includegraphics[width=0.5\textwidth]{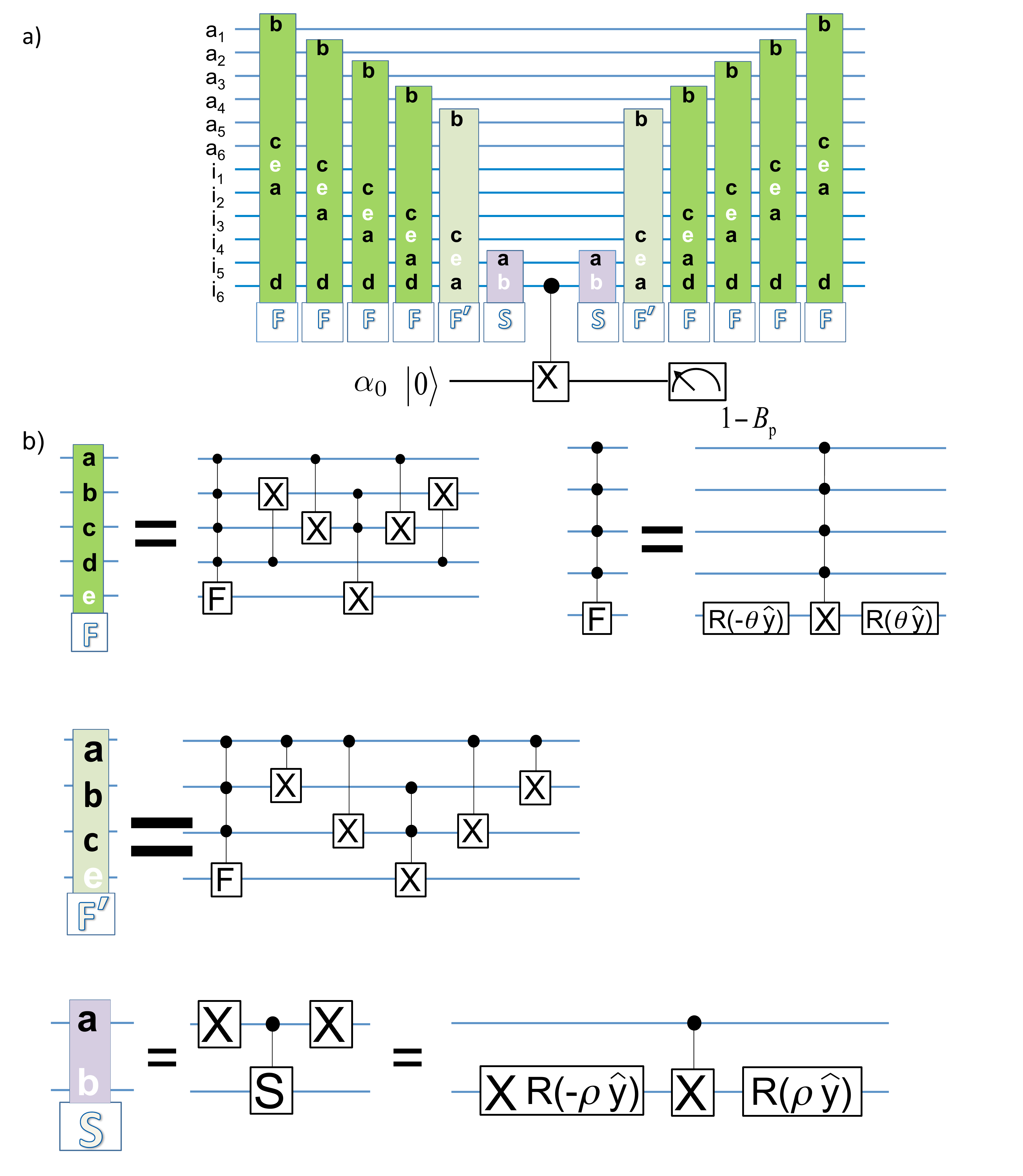}
	\caption{Figure reproduced with permission from Ref.~\onlinecite{BDV}: Quantum circuit for the plaquette reduction method. a) Full circuit for the plaquette reduction method to calculate the value of the plaquette operator $B_{p}$. The numbering of the qubits is that of Fig.~\ref{fig:LW}b). The individual gates of the circuit are detailed in b). b) Each element of the circuit in a) is reduced to  $X$-gates, $S$-gates, single qubit rotations $R(\rho\,\hat{y})$ by an angle $\rho$ along the $y$-axis, controlled-$X$ gates, controlled-$S$ gates, and Toffoli gates.}
	\label{fig:Reduction}
\end{figure}

Recently, Ref.~\onlinecite{BDV} has shown how to explicitly construct quantum circuits that measure plaquette and vertex operators of the Fibonacci Levin-Wen model; this is required to measure the error syndrome of $\mathcal{F}$ and to decide how to perform error correction. Here we go one step further and determine the optimal qubit-coupler architecture to realize those quantum circuits.
It is not the goal of the present work to review in detail how vertex and plaquette quantum circuits are constructed. But these circuits indicate which qubits must be coupled and this indicates the binary linear program of Sec.~\ref{sec:1} that is to be solved to obtain the optimal architecture. For the sake of completeness in Fig.~\ref{fig:Reduction} we reproduce the circuit of Ref.~\onlinecite{BDV} for the \emph{plaquette reduction} method.

We note here that ancillary qubits are needed to perform non-demolition measurements of $Q_{v}$ and $B_{p}$. According to the plaquette reduction method of Ref.~\onlinecite{BDV}, in Fig.~\ref{fig:LW} the ancillary qubit $\alpha_{0}$ is used to measure $B_{p}$, while the ancillary qubits $\alpha_{1-6}$ are used to measure the six vertex operators $Q_{v}$.

Here we choose the cost function $\mathcal{C}_{\mathcal{S}}$ that measures the geometric length of the TLR corresponding to $\mathcal{S}=\{i_{1}, i_{2},\ldots, i_{m}\}$,
\begin{equation}
\mathcal{C}_\mathcal{S}=\min_{\sigma}\left\{\sum_{k=2}^{m}\vert {\bf x}_{i_{\sigma(k)}}-{\bf x}_{i_{\sigma(k-1)}}\vert\right\}\,,
\end{equation}
where $\sigma$ is a permutation of $m$ elements.

Said differently, $\mathcal{C}_{\mathcal{S}}$ is the geometric length of the shortest path going through all the TQs specified in the string $\mathcal{S}$. In this work we thus look for the TLR scheme that minimizes the total length of the TLR wires.

Following the plaquette reduction method of Ref.~\onlinecite{BDV} and using the notation of Fig.~\ref{fig:LW}, we present in Table \ref{tab:1} the set $\mathcal{P}_{\text{reduction}}$ of two-qubit couplings that are necessary to measure the six vertex operators and the single plaquette operator of Fig.~\ref{fig:LW}.
\begin{table}[h!]
\centering
\begin{tabular}{|c|c|}
\hline\hline
Primary Qubit & Qubits to which the primary qubit couples \\
\hline
$\alpha_{0}$ & $i_{6}$\\
$a_{1}$ & $a_{6},i_{1},i_{2},i_{6},\alpha_{1}$\\
$a_{2}$ & $i_{1},i_{2},i_{3},i_{6},\alpha_{2}$\\
$a_{3}$ & $i_{2},i_{3},i_{4},i_{6},\alpha_{3}$\\
$a_{4}$ & $i_{3},i_{4},i_{5},i_{6},\alpha_{4}$\\
$a_{5}$ & $i_{4},i_{5},i_{6},\alpha_{5}$\\
$a_{6}$ & $a_{1},i_{1},i_{2},i_{6},\alpha_{6}$\\
$i_{1}$ & $a_{1},a_{2},a_{6},i_{2},i_{3},i_{6},\alpha_{1},\alpha_{6}$\\
$i_{2}$ & $a_{1},a_{2},a_{6},i_{1},i_{3},i_{4},i_{6},\alpha_{1},\alpha_{2}$\\
$i_{3}$ & $a_{2},a_{3},a_{4},i_{1},i_{2},i_{4},i_{5},i_{6},\alpha_{2},\alpha_{3}$\\
$i_{4}$ & $a_{3},a_{4},a_{5},i_{2},i_{3},i_{5},i_{6},\alpha_{3},\alpha_{4}$\\
$i_{5}$ & $a_{4},a_{5},i_{3},i_{4},i_{6},\alpha_{4},\alpha_{5}$\\
$i_{6}$ & $\alpha_{0},a_{1},a_{2},a_{3},a_{4},a_{5},a_{6},i_{1},i_{2},i_{3},i_{4},i_{5},\alpha_{5},\alpha_{6}$\\
$\alpha_{1}$ & $a_{1},i_{1},i_{2}$\\
$\alpha_{2}$ & $a_{2},i_{2},i_{3}$\\
$\alpha_{3}$ & $a_{3},i_{3},i_{4}$\\
$\alpha_{4}$ & $a_{4},i_{4},i_{5}$\\
$\alpha_{5}$ & $a_{5},i_{5},i_{6}$\\
$\alpha_{6}$ & $a_{6},i_{1},i_{6}$\\
\hline
\hline
\end{tabular}
\caption{We list the set $\mathcal{P}_{\text{reduction}}$ of all the two-qubit couplings required to measure the plaquette $p$ and the six vertex operators of Fig.~\ref{fig:LW}, following the plaquette reduction method of Ref.~\onlinecite{BDV}. The data and ancillary qubits are labeled according to the notation of Fig.~\ref{fig:LW}.}
\label{tab:1}
\end{table}

\begin{figure}[h!]
	\centering
		\includegraphics[width=0.4\textwidth]{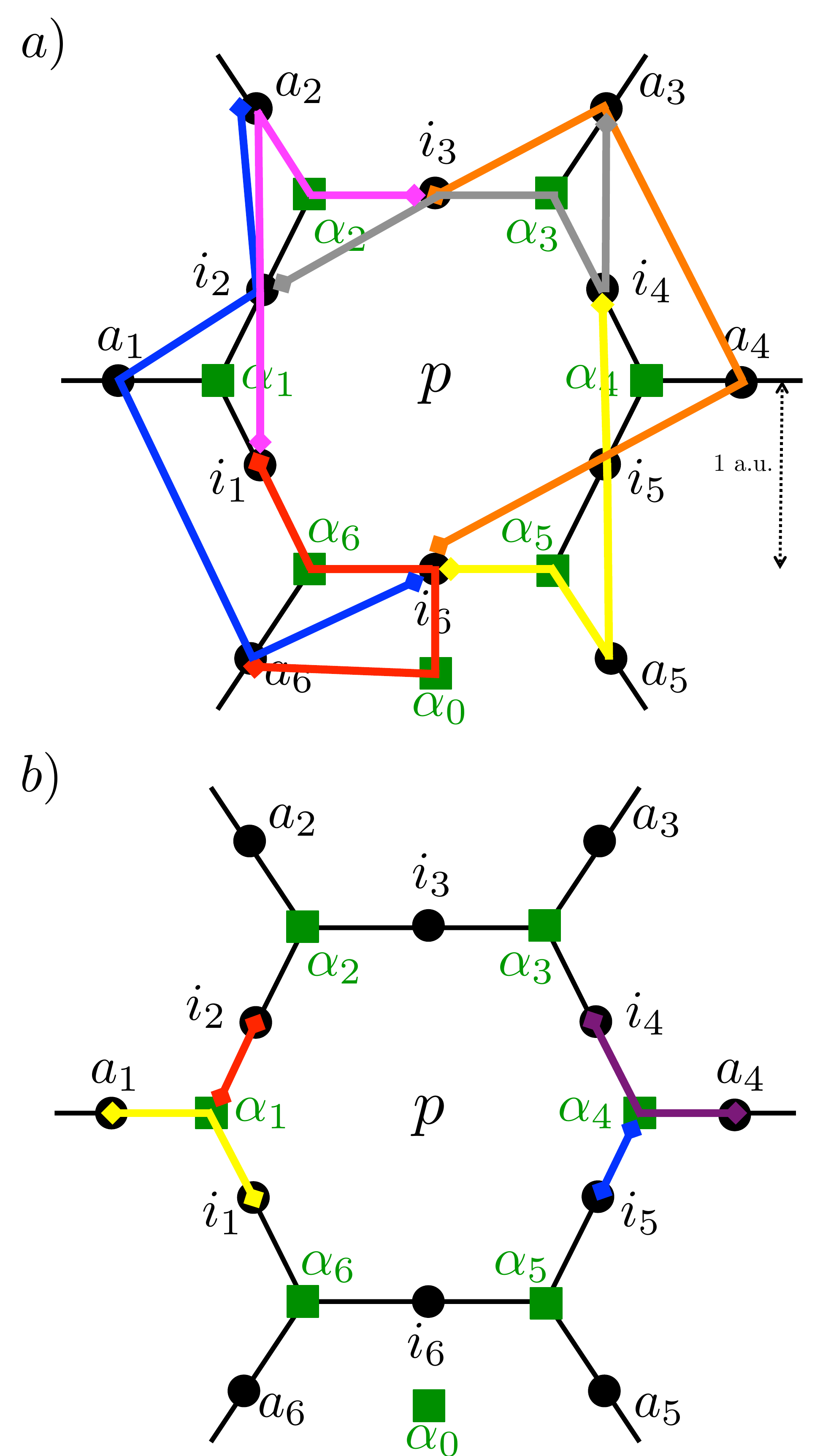}
	\caption{Pictorial representation of the optimal architecture of Table \ref{tab:2}. In $a)$ we show the TLRs that contain five TQs. The length of the arbitrary unit (a.u.) is depicted by the dashed arrow. In $b)$ we show the TLRs that contain three and two TQs.}
	\label{fig:LW2}
\end{figure}

Having in hand $\mathcal{P}_{\text{reduction}}$, we can solve the binary linear program of Sec.~\ref{sec:1} and determine the optimal TLR scheme. The result is summarized in Table \ref{tab:2} and a pictorial representation is given in Fig.~\ref{fig:LW2}.
\begin{table}[h!]\label{tab:2}
\centering
\begin{tabular}{|c|c|}
\hline
\hline
Length of the TLR wire & Qubits contained inside the TLR wire\\
\hline
2.52 a.u. & $a_{6},\alpha_{0},i_{6},\alpha_{6},i_{1}$\\
4.73 & $i_{6},a_{6},a_{1},i_{2},a_{2}$\\
3.15 & $i_{1},i_{2},a_{2},\alpha_{2},i_{3}$\\
4.73 & $i_{2},i_{3},\alpha_{3},i_{4},a_{3}$\\
3.15 & $i_{3},a_{3},a_{4},i_{5},i_{6}$\\
3.15 & $i_{6},\alpha_{5},a_{5},i_{5},i_{4}$\\
\hline
1.15 & $i_{1},\alpha_{1},a_{1}$\\
1.15 & $i_{4},\alpha_{4},a_{4}$\\
\hline
0.58 & $i_{2},\alpha_{1}$\\
0.58 & $i_{5},\alpha_{4}$\\
\hline
\hline
\end{tabular}
\caption{Information about the optimal TLR scheme obtained by solving the binary linear program for the measurement of the plaquette operator $B_{p}$ and the six vertex operators $Q_{v}$ of Fig.~\ref{fig:LW}, following the plaquette reduction method of Ref.~\onlinecite{BDV}. In particular, all the two-qubit couplings of $\mathcal{P}_{\text{reduction}}$ in Table \ref{tab:1} are realized: there are no more than five TQs in each TLR, and each TQ couples to maximally four TLRs. A pictorial representation of this TLR scheme and of the arbitrary unit (a.u.) is given in Fig.~\ref{fig:LW2}.}
\label{tab:2}
\end{table}

\begin{figure}[h!]
	\centering
		\includegraphics[width=0.4\textwidth]{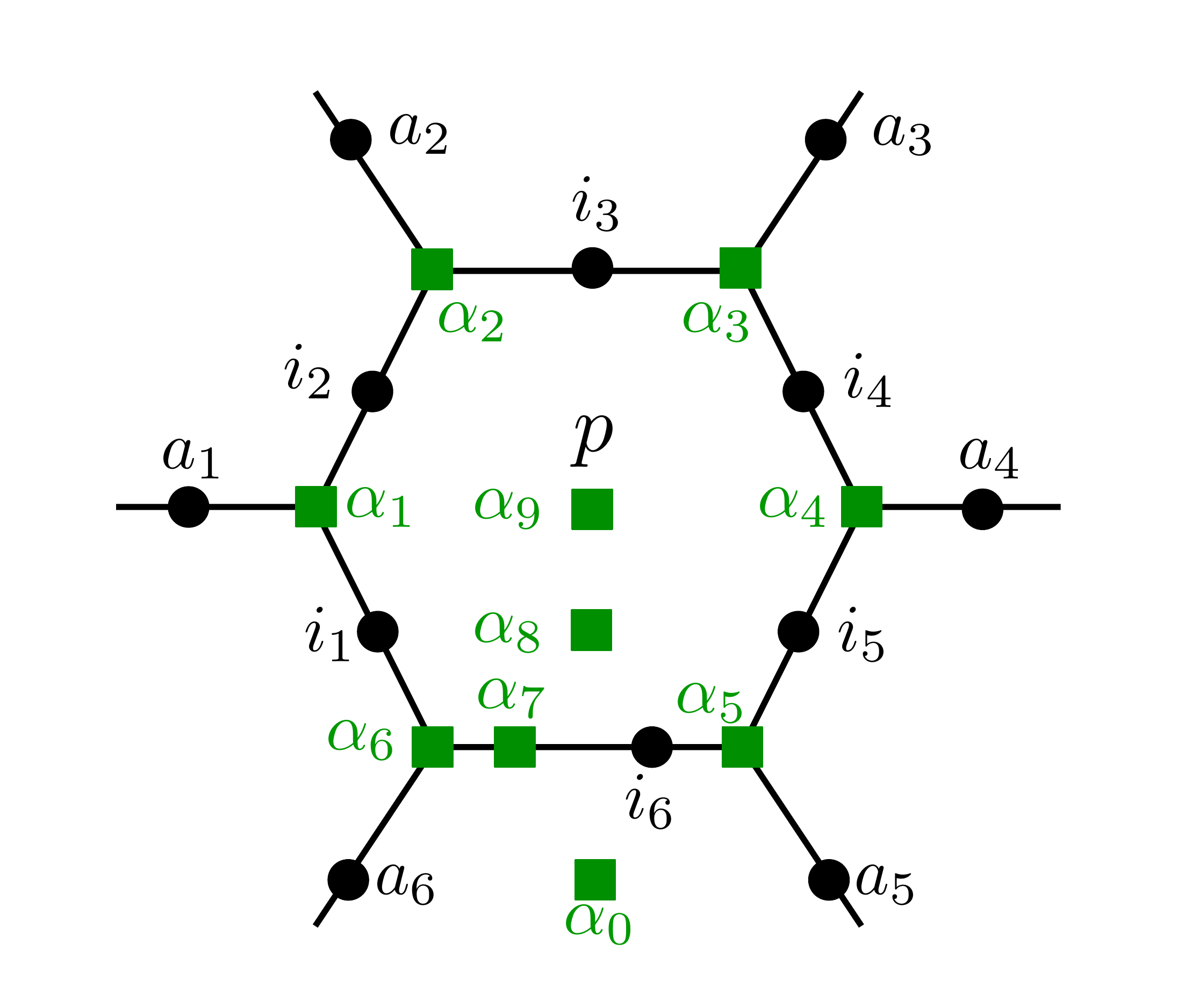}
	\caption{The data qubits of the Fibonacci Levin-Wen model are represented by black dots.  This qubit layout is appropriate for the plaquette swapping method of Ref.~\onlinecite{Weibo}. The ancillary qubits, necessary to perform non-demolition measurements of plaquette and vertex operators, are represented by green squares.}
	\label{fig:LW3}
\end{figure}

For completeness, we also investigate the \emph{plaquette swapping} method of Ref.~\onlinecite{Weibo} to measure plaquette operators. In this case, more ancillary qubits are required, see Fig.~\ref{fig:LW3}.
\begin{table}[h!]
\centering
\begin{tabular}{|c|c|}
\hline\hline
Primary Qubit & Qubits to which the primary qubit couples \\
\hline
$\alpha_{0}$ & $\alpha_{9}$\\
$a_{1}$ & $i_{1},i_{2},\alpha_{1},\alpha_{7},\alpha_{9}$\\
$a_{2}$ & $i_{2},i_{3},\alpha_{2},\alpha_{7},\alpha_{9}$\\
$a_{3}$ & $i_{3},i_{4},\alpha_{3},\alpha_{7},\alpha_{9}$\\
$a_{4}$ & $i_{4},i_{5},\alpha_{4},\alpha_{7},\alpha_{9}$\\
$a_{5}$ & $i_{5},i_{6},\alpha_{5},\alpha_{7},\alpha_{9}$\\
$a_{6}$ & $i_{1},i_{6},\alpha_{6},\alpha_{7},\alpha_{8},\alpha_{9}$\\
$i_{1}$ & $a_{1},a_{6},i_{2},i_{6},\alpha_{1},\alpha_{6},\alpha_{7},\alpha_{8},\alpha_{9}$\\
$i_{2}$ & $a_{1},a_{2},i_{1},i_{3},\alpha_{1},\alpha_{2},\alpha_{7},\alpha_{9}$\\
$i_{3}$ & $a_{2},a_{3},i_{2},i_{4},\alpha_{2},\alpha_{3},\alpha_{7},\alpha_{9}$\\
$i_{4}$ & $a_{3},a_{4},i_{3},i_{5},\alpha_{3},\alpha_{4},\alpha_{7},\alpha_{9}$\\
$i_{5}$ & $a_{4},a_{5},i_{4},i_{6},\alpha_{4},\alpha_{5},\alpha_{7},\alpha_{9}$\\
$i_{6}$ & $a_{5},a_{6},i_{1},i_{5},\alpha_{5},\alpha_{6},\alpha_{7},\alpha_{8},\alpha_{9}$\\
$\alpha_{1}$ & $a_{1},i_{1},i_{2}$\\
$\alpha_{2}$ & $a_{2},i_{2},i_{3}$\\
$\alpha_{3}$ & $a_{3},i_{3},i_{4}$\\
$\alpha_{4}$ & $a_{4},i_{4},i_{5}$\\
$\alpha_{5}$ & $a_{5},i_{5},i_{6}$\\
$\alpha_{6}$ & $a_{6},i_{1},i_{6}$\\
$\alpha_{7}$ & $a_{1},a_{2},a_{3},a_{4},a_{5},a_{6},i_{1},i_{2},i_{3},i_{4},i_{5},i_{6},\alpha_{8},\alpha_{9}$\\
$\alpha_{8}$ & $a_{6},i_{1},i_{6},\alpha_{7},\alpha_{9}$\\
$\alpha_{9}$ & $\alpha_{0},a_{1},a_{2},a_{3},a_{4},a_{5},a_{6},i_{1},i_{2},i_{3},i_{4},i_{5},i_{6},\alpha_{7},\alpha_{8}$\\
\hline
\hline
\end{tabular}
\caption{The set $\mathcal{P}_{\text{swapping}}$ of all the two-qubit couplings required to measure the plaquette $p$ and the six vertex operators of Fig.~\ref{fig:LW3}, following the plaquette swapping method of Ref.~\onlinecite{Weibo}. The data and ancillary qubits are labeled according to the notation of Fig.~\ref{fig:LW3}.}
\label{tab:3}
\end{table}
The set $\mathcal{P}_{\text{swapping}}$ of two-qubit couplings required by the plaquette swapping method is summarized in Table \ref{tab:3}. Again, we solve the binary linear program and find the optimal architecture of Table \ref{tab:4};  here we have again chosen $m=p=5$.  As the pictorial representation would be too crowded, we refrain from drawing the TLRs corresponding to Table \ref{tab:4}.
\begin{table}[h!]
\centering
\begin{tabular}{|c|c|}
\hline\hline
Length of the TLR wire & Qubits contained inside the TLR wire \\
\hline
2.52 a.u. & $\alpha_{9},\alpha_{8},i_{6},\alpha_{0},a_{6}$\\
2.84 & $a_{1},i_{1},\alpha_{7},\alpha_{8},\alpha_{9}$\\
4.04 & $\alpha_{7},\alpha_{9},i_{4},i_{3},a_{3}$\\
4.30 & $\alpha_{9},\alpha_{7},a_{5},i_{5},a_{4}$\\
2.44 & $i_{1},\alpha_{7},\alpha_{6},i_{6},a_{6}$\\
\hline
2.15 & $i_{2},\alpha_{1},i_{1},a_{1}$\\
2.15 & $i_{3},\alpha_{2},i_{2},a_{2}$\\
3.04 & $\alpha_{7},\alpha_{9},i_{2},a_{2}$\\
2.15 & $i_{5},\alpha_{4},i_{4},a_{4}$\\
1.63 & $i_{5},\alpha_{5},i_{6},a_{5}$\\
\hline
1.15 & $i_{4},\alpha_{3},a_{3}$\\
\hline
0.58 & $i_{3},\alpha_{3}$\\
\hline
\hline
\end{tabular}
\caption{The optimal TLR scheme obtained by solving the binary linear program for the measurement of the plaquette operator $B_{p}$ and the six vertex operators $Q_{v}$ of Fig.~\ref{fig:LW3}, following the plaquette swapping method of Ref.~\onlinecite{Weibo}. In particular, all the two-qubit couplings of $\mathcal{P}_{\text{swapping}}$ in Table \ref{tab:3} are realized: there are no more than five TQs in each TLR, and each TQ couples to maximally four TLRs.}
\label{tab:4}
\end{table}
\begin{figure}[h!]
	\centering
		\includegraphics[width=0.45\textwidth]{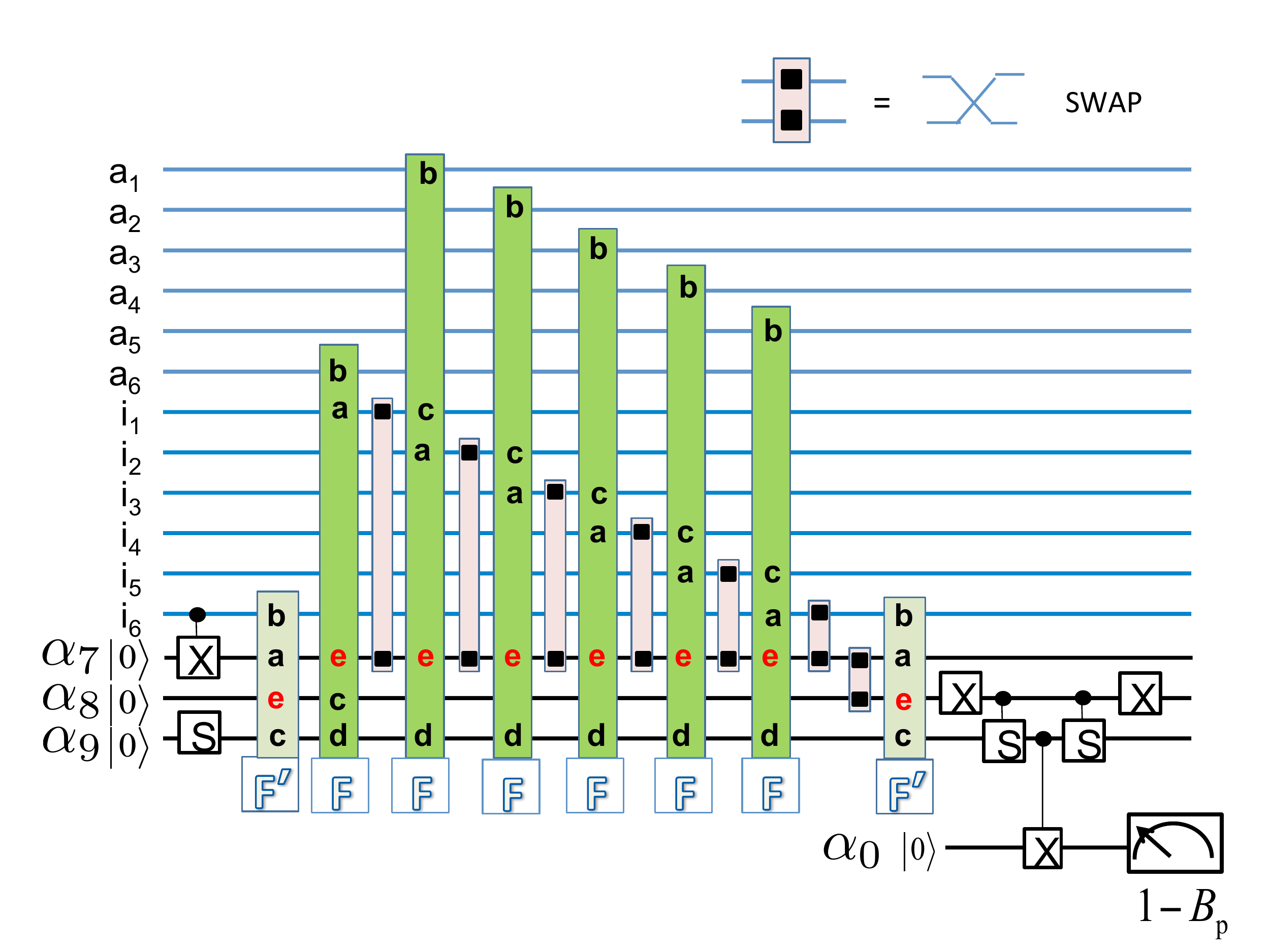}
	\caption{Figure reproduced with permission from Ref.\onlinecite{Weibo}. Quantum circuit for the plaquette swapping method of Ref.~\onlinecite{Weibo}. The qubits labeling is the one of Fig.~\ref{fig:LW3}. The individual elements of the circuits can be found in Fig.~\ref{fig:Reduction}.}
	\label{fig:Swapping}
\end{figure}

\subsubsection{Scaling}
\begin{figure}[h!]
	\centering
		\includegraphics[width=0.4\textwidth]{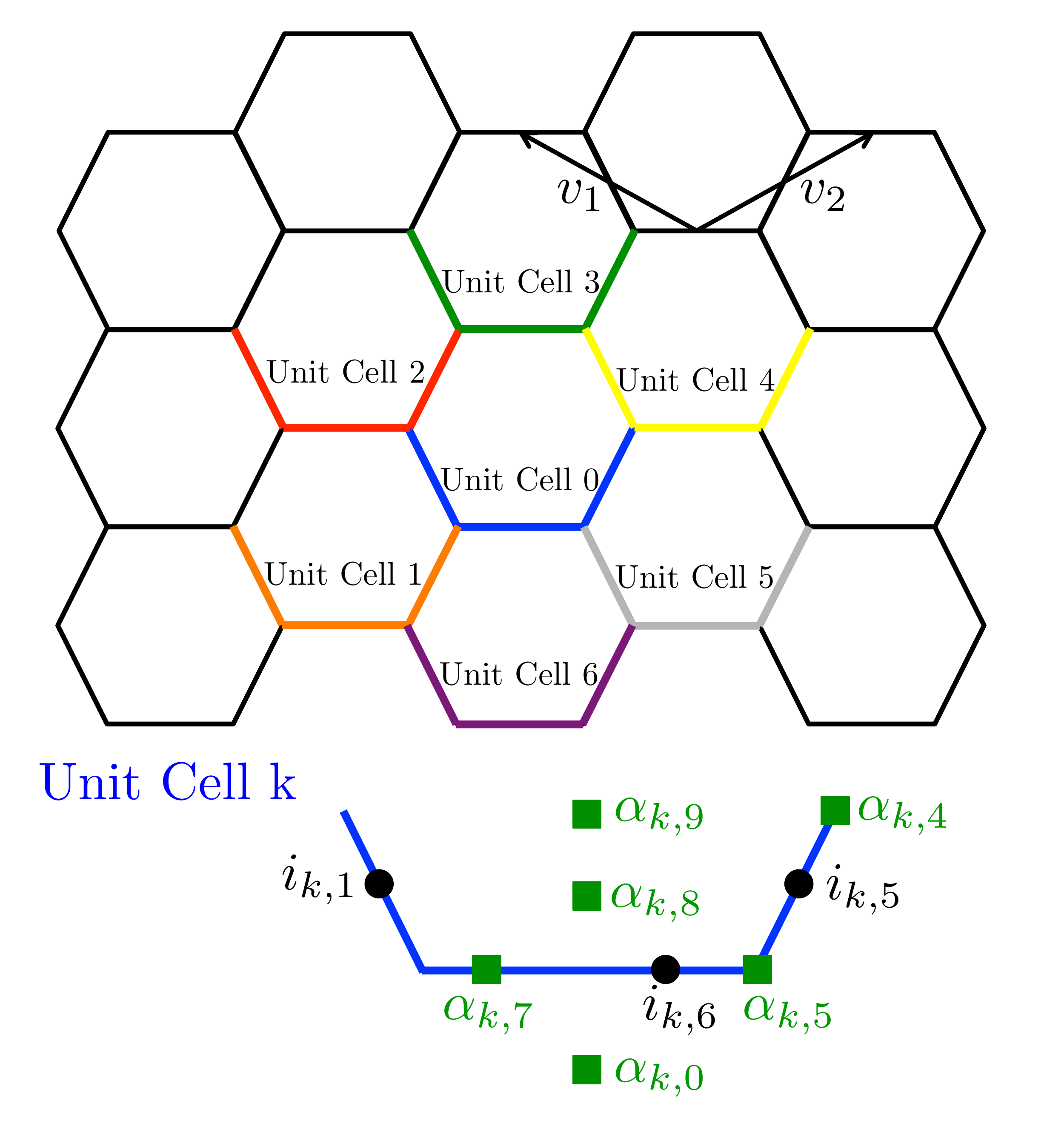}
	\caption{Trivalent lattice on which the Levin-Wen model is defined. The lattice is obtained by translating the unit cell 0 by the unit vectors ${\bf v}_{1}$ and ${\bf v}_{2}$. For example, the unit cell 1 is obtained by translating unit cell 0 by $-{\bf v}_{2}$. For the sake of clarity, we do not represent all the qubits of the lattice. Instead, we draw and label the data (black dots) and ancillary (green squares)  qubits of a given unit cell $k$. This qubit layout is appropriate for the plaquette swapping method.\cite{Weibo}}
	\label{fig:UnitCell}
\end{figure}

While binary linear programs can be solved rapidly for a small number of qubits, as is the case for the 12 data qubits of Figs.~\ref{fig:LW} and \ref{fig:LW3}, the problem becomes rapidly unsolvable when we increase the number of qubits. This seems to be problematic as one wants to find the optimal architecture for a large Levin-Wen model and not only for a single plaquette.  Fortunately, most of the time there is a lot of redundancy in the problem in the sense that a fundamental circuit unit can be identified and translated over the whole lattice. In fact, if one wants for example to measure all the vertex and plaquette operators of a large Fibonacci Levin-Wen model, the circuit will look the same around any plaquette of the lattice. In such a scenario, it is possible to identify a small number of qubits that we couple optimally and that we translate to cover the whole lattice.  The aim of this section is thus to introduce a simple method to optimally solve a given unit cell of the model that can be scaled up by simple translation to build a large two-dimensional lattice, see Fig.~\ref{fig:UnitCell}. For the sake of simplicity, we just focus here on the plaquette swapping method of Ref.~\onlinecite{Weibo}.
In Table \ref{tab:5} we present the set $\mathcal{P}_{\text{swapping}}^{0}$ of two-qubit couplings  that are required between the qubits of unit cell 0, see Fig.~\ref{fig:UnitCell}, and the remaining qubits of the lattice.
\begin{widetext}
\begin{center}
\begin{table}[h!]
\centering
\begin{tabular}{|c|c|}
\hline\hline
Qubit $q$ in unit cell $0$ & Qubits to which $q$ couples  \\
\hline
$i_{0,1}$ & $\alpha_{0,7},i_{0,6},\alpha_{0,8},\alpha_{0,9},i_{1,5},\alpha_{1,4},i_{1,6},\alpha_{1,9},\alpha_{2,7},i_{2,5},\alpha_{2,5},i_{2,6},\alpha_{2,9},i_{6,6},\alpha_{6,9}$\\
$i_{0,5}$ &$\alpha_{0,7},\alpha_{0,5},\alpha_{0,4},i_{0,6},\alpha_{0,9},\alpha_{4,7},i_{4,1},i_{4,6},\alpha_{4,7},\alpha_{4,9},i_{5,1},i_{5,6},\alpha_{5,9},i_{6,6},\alpha_{6,9}$\\
$i_{0,6}$ & $\alpha_{0,7},i_{0,1},i_{0,5},\alpha_{0,8},\alpha_{0,9},i_{1,5},\alpha_{2,7},i_{2,5},\alpha_{3,7},i_{3,1},i_{3,5},\alpha_{4,7},i_{4,1},i_{5,1}$\\
$\alpha_{0,0}$ & $\alpha_{0,9}$\\
$\alpha_{0,4}$ & $i_{0,5},\alpha_{4,7},i_{4,1}$\\
$\alpha_{0,5}$ & $\alpha_{0,7},i_{0,5},i_{5,1}$\\
$\alpha_{0,7}$ &$i_{0,1},i_{0,5},\alpha_{0,5},i_{0,6},\alpha_{0,8},\alpha_{0,9},i_{1,5},\alpha_{1,4},i_{1,6},\alpha_{1,9},i_{5,1},i_{5,6},\alpha_{5,9},i_{6,6},\alpha_{6,9}$\\
$\alpha_{0,8}$ & $\alpha_{0,7},i_{0,1},i_{0,6},\alpha_{0,9},i_{1,5},$\\
$\alpha_{0,9}$ & $\alpha_{0,7},i_{0,1},i_{0,5},\alpha_{0,0},i_{0,6},\alpha_{0,8},i_{1,5},\alpha_{2,7},i_{2,5},\alpha_{3,7},i_{3,1},i_{3,5},\alpha_{4,7},i_{4,1},i_{5,1}$\\
\hline
4 & $\alpha_{0,7},i_{0,1},\alpha_{0,8},\alpha_{1,4}$\\
\hline
3 & $i_{0,5},\alpha_{0,5},\alpha_{0,4}$\\
\hline
\hline
\end{tabular}
\caption{The set $\mathcal{P}_{\text{swapping}}^{0}$ of two-qubit connections between qubits in unit cell 0 of Fig.~\ref{fig:UnitCell} and the remaining qubits of the lattice, following the plaquette swapping method of Ref.~\onlinecite{Weibo}.}
\label{tab:5}
\end{table}
\end{center}
\end{widetext}
If one would now straightforwardly solve the binary linear program for the unit cell, as we did in Sec.~\ref{sec:FLW}, then one would encounter the problem of \emph{equivalent connections},  i.e., connections that are doubled due to the shifting of the unit cell. As an explicit example, let us consider the connection between qubits $i_{0,6}$ and $i_{2,5}$ as well as the connection between qubits $i_{0,5}$  and $i_{5,6}$, see Fig.~\ref{fig:UnitCell}. It is straightforward to see that after translating unit cell 0 onto unit cell 5, a TLR will be doubled.
In order to avoid such doublings, one needs to slightly modify the algorithm as follows.

Consider a given unit cell $0$ and two distinct strings $\mathcal{S}_{1}$ and $\mathcal{S}_{2}$ that each contains \emph{at least one} site inside unit cell $0$. We say that $\mathcal{S}_{1}=\{i_{1},i_{2},\ldots, i_{m}\}$ and $\mathcal{S}_{2}=\{j_{1},j_{2},\ldots, j_{m}\}$ are equivalent if $\forall k\in[1,m]$ $\exists\, \ell\in[1,m]$ such that
\begin{equation}
{\bf x}_{i_{k}}={\bf x}_{j_{\ell}}+\lambda_{1}{\bf v}_{1}+\lambda_{2}{\bf v}_{2}\,,
\end{equation}
where $\lambda_{1,2}\in\mathbb{Z}$ and ${\bf v}_{1,2}$ are basis vectors of the lattice, see Fig.~\ref{fig:UnitCell}.  If a TLR scheme possesses a TLR hosting the qubits along $\mathcal{S}_{1}$ and another TLR hosting  the qubits along $\mathcal{S}_{2}$, it is clear that this will not be optimal. Indeed, when we translate unit cell $0$ by the vector $\lambda_{1}{\bf v}_{1}+\lambda_{2}{\bf v}_{2}$ and the associated TLRs, to cover the whole lattice, then some TLRs will be doubled. Having set these definitions, we present the steps that we follow to find the optimal scalable TLR without doubled TLRs.

\begin{itemize}

\item Consider the set $\mathcal{P}^{0}_{\text{swapping}}$ of two-qubit couplings that contains at least one qubit in the unit cell 0. 
\item Define $\mathfrak{W}$ as the set of TLR schemes that include all TLRs that are not superfluous with respect to $\mathcal{P}_{\text{swapping}}^{0}$ and all their equivalent TLRs.
\item Out of every set of equivalent TLRs, choose one unique representative TLR. For each TLR scheme $W_{m}\in\mathfrak{W}$, define an associated TLR scheme $B_{m}$. This scheme $B_{m}$ includes the same TLRs as $W_{m}$ but contains the following TLRs: All TLRs that do not have an equivalent TLR and are contained in $W_{m}$ as well as all representatives for which $W_{m}$ contains at least one equivalent TLR. We call this new set of TLR schemes $\mathfrak{B}$.
\item For each TLR scheme $B_{m}\in\mathfrak{B}$, define a new TLR scheme $V_{m}$ that includes the same TLRs as $B_{m}$ and contains all the TLRs that are contained in $B_{m}$ as well as all equivalent TLRs.
\item Perform the linear optimization over $\mathfrak{B}$ to find a TLR scheme $B_{m}$ that minimizes the cost $C(B_{m})^{t}\cdot B_{m}$ such that
\begin{enumerate}
\item $V_{m}$ realizes every two-qubit coupling of $\mathcal{P}^{0}_{\text{swapping}}$.
\item For all $i_{j}\in\{1,2,\ldots,N\}$, \begin{equation}\sum_{\mathcal{S}\in\mathfrak{S}_{m}\,\vert\, i_{j}\in\mathcal{S}}\kappa_{\mathcal{S}}\leqslant p\,.\end{equation}
\end{enumerate}

\end{itemize}

\begin{figure}[h!]
	\centering
		\includegraphics[width=0.5\textwidth]{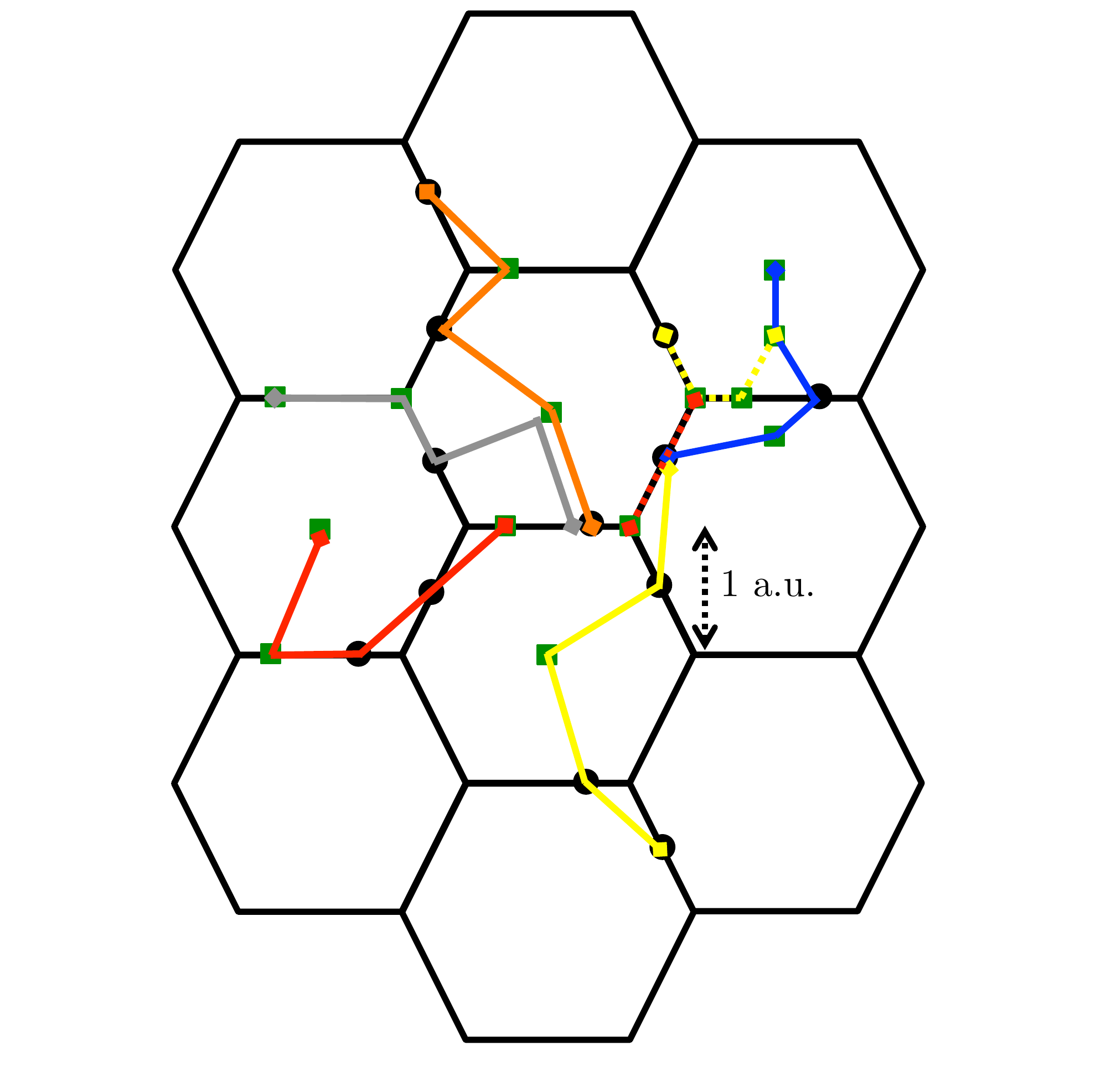}
	\caption{Pictorial representation of the optimal scalable solution of Table \ref{tab:6}. The TLRs couple qubits inside unit cell 0 to the rest of the lattice. This architecture can be translated to cover the whole lattice without generating doubled  TLRs. The TLRs are represented by solid and dashed lines for clarity when they traverse the same path. The squares denote the starting and ending points of TLRs. Note that the four qubit TLR, that contains qubits $\alpha_{0,8}, \alpha_{0,7}, \alpha_{1,4}, i_{0,1}$ has not be drawn, but instead we have drawn it translated (yellow, dashed) for the sake of a clear figure.}
	\label{fig:UnitCellSolution}
\end{figure}

Following the above algorithm, we find the scalable optimal architectures presented in Table \ref{tab:6} and Fig.~\ref{fig:UnitCellSolution}.
\begin{table}[h!]
\centering
\begin{tabular}{|c|c|}
\hline\hline
Length of the TLR & Qubits contained inside the TLR wire  \\
\hline
3.57 a.u. & $i_{0,6},\alpha_{0,9},i_{2,5},\alpha_{3,7},i_{3,1}$\\
3.15 & $\alpha_{1,9},\alpha_{1,7},i_{1,6},i_{1,5},\alpha_{0,7}$\\
3.48 & $\alpha_{2,7},\alpha_{2,5},i_{0,1},\alpha_{0,9},i_{0,6}$\\
2.52 & $i_{0,5},\alpha_{4,0},i_{4,6},\alpha_{4,8},\alpha_{4,9}$\\
3.80 & $i_{17,1},i_{6,6},\alpha_{6,9},i_{5,1},i_{0,5}$\\
\hline
1.44 & $\alpha_{0,8},\alpha_{0,7},\alpha_{1,4},i_{0,1}$\\
\hline
1.15 & $\alpha_{0,4},i_{0,5},\alpha_{0,5}$\\
\hline
\hline
\end{tabular}
\caption{Optimal TLR scheme for the set of couplings $\mathcal{P}_{\text{swpapping}}^{0}$, i.e., for couplings between qubits in the unit cell 0 and the rest of the lattice. By translating this TLR scheme to the remaining unit cells of the lattice, one obtains an optimal TLR scheme for the entire lattice that avoids doubled TLRs.}
\label{tab:6}
\end{table}

\subsection{Surface Code}\label{sec:SC}
The surface code \cite{TerhalReview} is a planar version of Kitaev's toric code \cite{KitaevToric}  and represents arguably the most promising quantum computing architecture. It is thus justified to  determine its optimal architecture using the simple formalism developed in this work,  in particular because experimental groups are nowadays starting to build small fragments of the surface code. 
\begin{figure}[h!]
	\centering
		\includegraphics[width=0.5\textwidth]{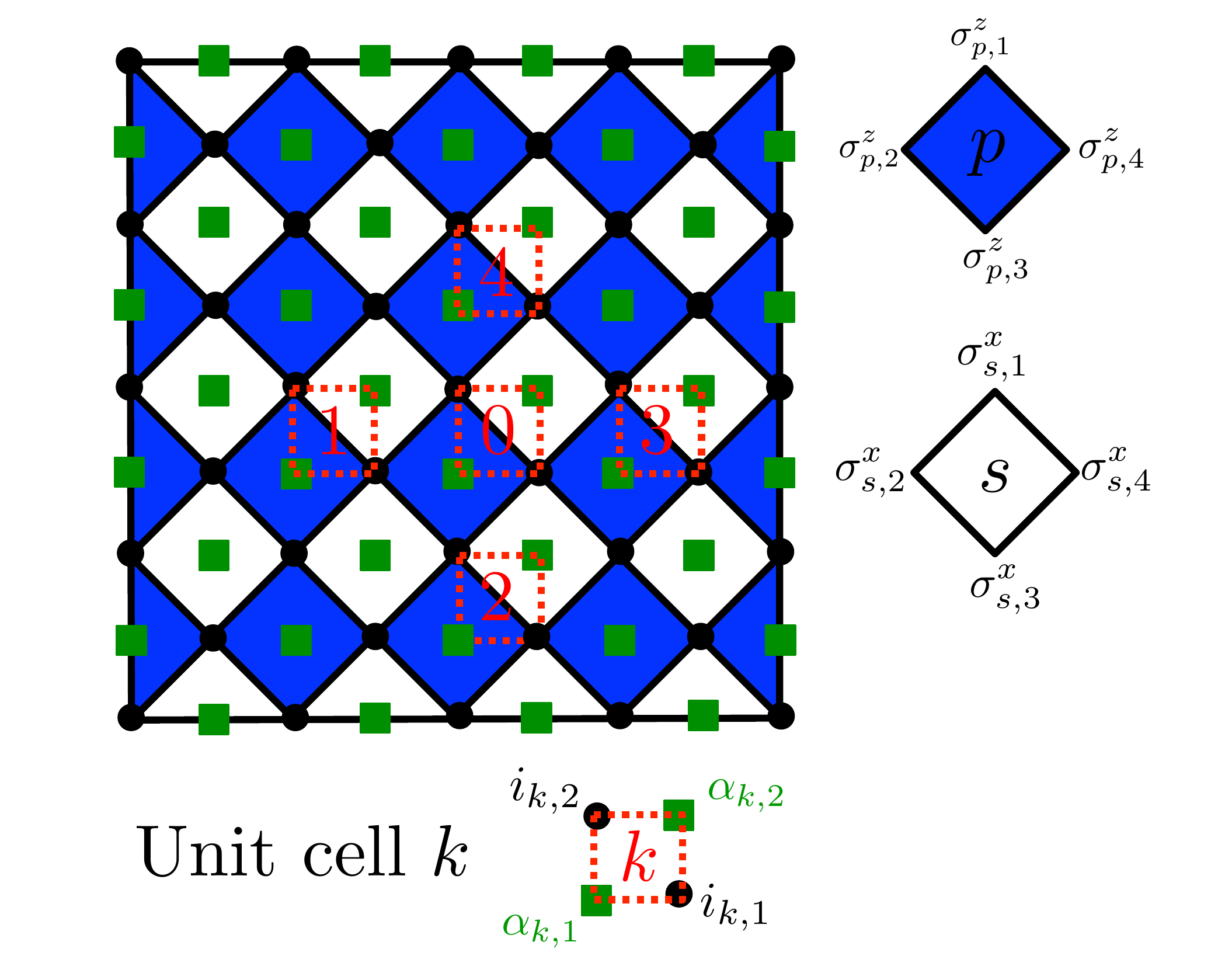}
	\caption{Pictorial representation of the surface code. Data qubits reside on  the vertices of the lattice and are here depicted by black dots. Products of data qubits around dark (light) squares correspond to plaquette (star) operators. The green squares are ancillary qubits necessary to measure plaquette and star operators in a non-demolition fashion. We choose the unit cell 0 (dashed square) to generate the whole lattice by translations. The two data qubits and the two ancillary qubits of a unit cell are labeled as shown in the figure. This choice of boundary conditions leads to a twofold degeneracy of the surface code $\mathcal{S}$ in Eq.~(\ref{eq:SC}).}
	\label{fig:SurfaceCode}
\end{figure}

Consider a square lattice with a spin-$1/2$ particle on each vertex, see Fig.~\ref{fig:SurfaceCode}. We define the star operators $A_{s}$ and plaquette operators $B_{p}$ of the surface code as
\begin{eqnarray}
A_{s}&=&\sigma_{s,1}^{x}\sigma_{s,2}^{x}\sigma_{s,3}^{x}\sigma_{s,4}^{x}\,,\\
B_{p}&=&\sigma_{p,1}^{x}\sigma_{p,2}^{x}\sigma_{p,3}^{x}\sigma_{p,4}^{x}\,,
\end{eqnarray}
where $s$ and $p$ label respectively light and dark squares of the lattice, see Fig.~\ref{fig:SurfaceCode}. Note that plaquette and star operators at the boundaries are products of three qubit operators and not four as is the case in the bulk. Similar to the Fibonacci code, we define the surface code $\mathcal{S}$ as
\begin{equation}\label{eq:SC}
\mathcal{S}=\{\vert\psi\rangle\, \vert\, A_{s}\vert\psi\rangle=B_{p}\vert\psi\rangle=\vert\psi\rangle, \forall p,s\}\,.
\end{equation}
With the boundary conditions represented in Fig.~\ref{fig:SurfaceCode}, the surface code is twofold degenerate and can thus encode a logical qubit.\cite{TerhalReview} Similar to the Fibonacci code, the surface code is a topological code and it is thus protected against local (static) perturbations. Its most striking property is its surprisingly high error threshold of about $1\%$, see Ref.~\onlinecite{TerhalReview} for a detailed review on this subject. 

Here we do not review the construction of quantum circuits to measure star and plaquette operators of the surface code, rather, using the notation of Fig.~\ref{fig:SurfaceCode}, in Table \ref{tab:7} we show the set $\mathcal{P}_{\text{surface}}^{0}$ of two-qubit couplings required to measure the eigenvalues of $A_{s}$ and $B_{p}$ in a scalable manner.\cite{Dennis} As was the case for the Fibonacci code, we need to introduce ancillary qubits to measure plaquette and star operators non-destructively.
\begin{table}[h!]
\centering
\begin{tabular}{|c|c|}
\hline\hline
Qubit $q$ in unit cell 0 & Qubits to which $q$ couples  \\
\hline
$i_{0,1}$ & $\alpha_{0,1},\alpha_{0,2},\alpha_{3,1},\alpha_{2,2}$\\
\hline
$\alpha_{0,1}$ & $i_{0,1},i_{0,2},i_{1,1},i_{2,2}$\\
\hline
$i_{0,2}$ & $\alpha_{0,1},\alpha_{0,2},\alpha_{1,2},\alpha_{4,1}$\\
\hline
$\alpha_{0,2}$ & $i_{0,1}, i_{0,2}, i_{4,1}, i_{3,2}$\\
\hline
\hline
\end{tabular}
\caption{The set $\mathcal{P}^{0}_{\text{surface}}$ of couplings between qubits of the unit cell 0 and the rest of the lattice to measure plaquette and star operators of the surface code, see Fig.~\ref{fig:SurfaceCode}.}
\label{tab:7}
\end{table}

\begin{figure}[h!]
	\centering
		\includegraphics[width=0.5\textwidth]{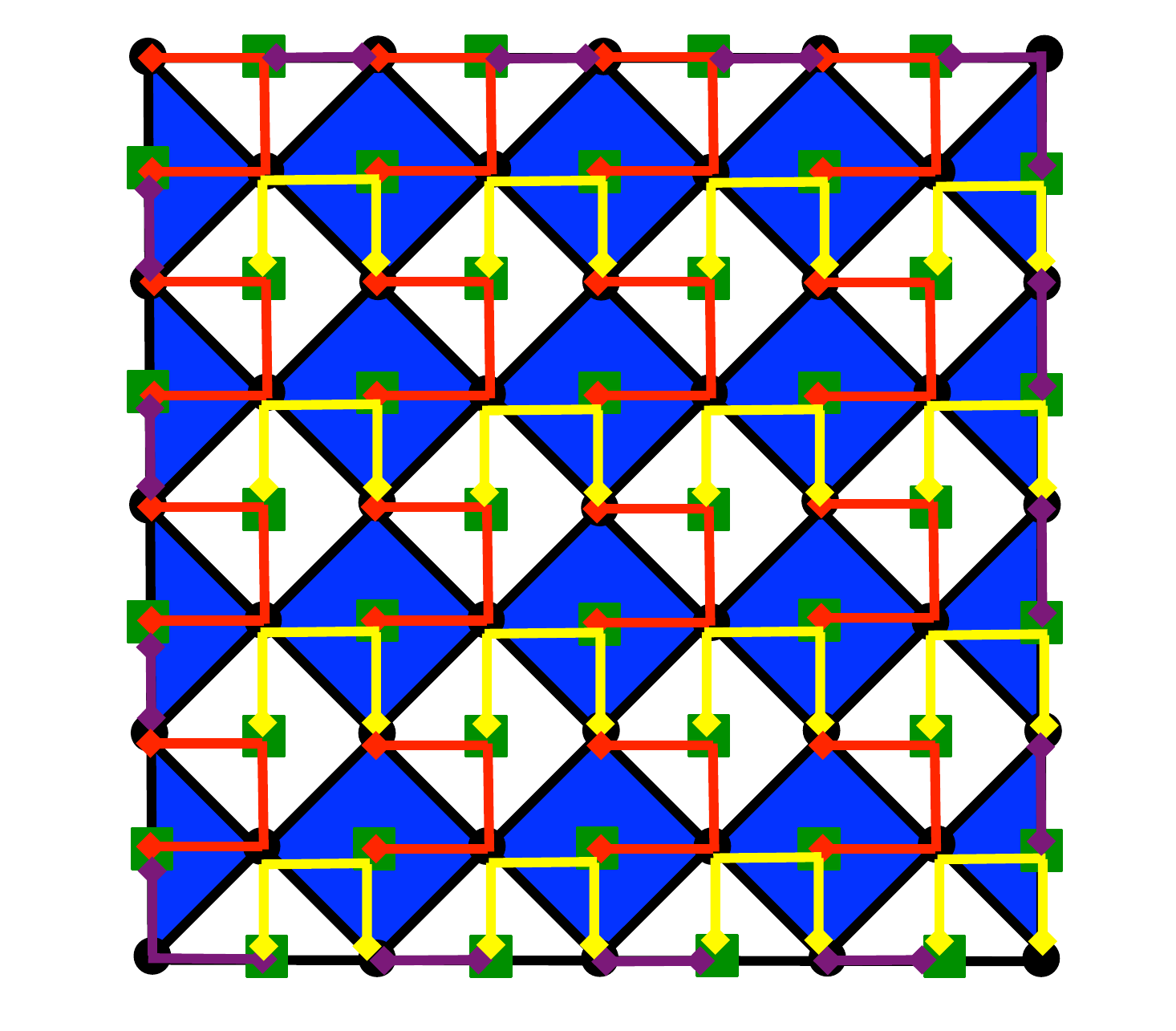}
	\caption{Optimal scalable TLR scheme for the surface code. The problem is first solved for the couplings between qubits of unit cell 0 and the rest of the lattice, and then we have translated the result to cover the whole lattice. This results is equivalent to the one originally proposed in Ref.~\onlinecite{DiVincenzo}. The TLRs are represented by solid lines and the squares denote the starting and ending points of TLRs. Note that rotating each individual four-qubit TLR by 90 degrees leads obviously to another optimal solution. The two-qubit TLRs on the edges and the three-qubit TLRs on the corners (purple) have been put by hand, since a full optimization solution of a smaller surface code lead to such a pattern of boundary two- and three-qubit TLRs. Indeed, it seems clear that that this solution at the boundaries remains optimal for a larger surface code.}
	\label{fig:SurfaceCode3}
\end{figure}

Solving the binary linear optimization problem with $m=5$,  we find the scalable optimal TLR scheme reported in Fig.~\ref{fig:SurfaceCode3}. The result is that each bulk TLR hosts four TQs and each bulk TQ is hosted by two TLRs. Interestingly, this result is the one originally proposed in Ref.~\onlinecite{DiVincenzo}, see also Ref.~\onlinecite{Ghosh}. We point out that the optimal architecture of Fig.~\ref{fig:SurfaceCode3} is clearly valid for the smallest possible surface code (consisting of 13 data qubits and 12 ancillary qubits) able to detect and correct a single physical qubit error. In fact, we have solved our optimization problem for this small surface code directly and obtained the solution of Fig.~\ref{fig:SurfaceCode3} with two- and three-qubit TLRs at the boundaries.

\subsubsection{Distance 5 surface code}
\begin{figure}[h!]
	\centering
		\includegraphics[width=0.45\textwidth]{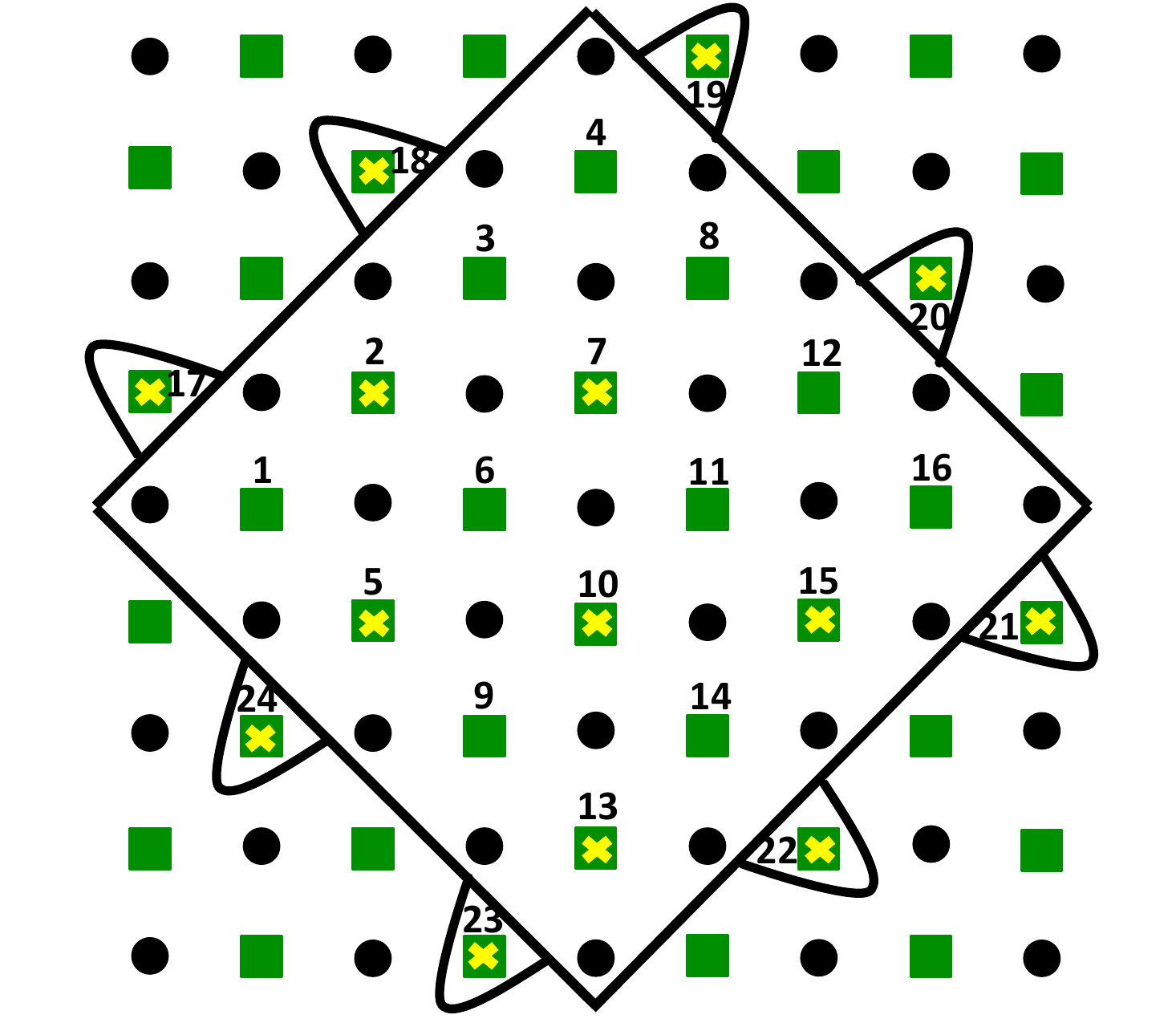}
	\caption{Pictorial representation of the rotated surface code. The black dots represent data qubits and the green squares ancillary qubits. The rotated surface code \cite{Horsman} contains all the qubits inside the solid square as well as the ancillary qubits surrounded by a black line. In this case, the rotated surface code contains 25 data qubits and 24 ancillary qubits; each ancillary qubit is used to measure exactly one stabilizer. However, one can slightly improve the resource needed by re-using ancillary qubits; we associate many of the ancillary qubits to more than one plaquette and star operator  measurements. We thus remove the ancillary qubits with a yellow cross. Here we list which qubits replace the crossed qubits in the syndrome-computation circuit: qubit 2 is replaced by qubit 3, qubit 5 is replaced by qubit 6, qubit 7 is replaced by qubit 8, qubit 10 is replaced by qubit 11, qubit 13 is replaced by qubit 14, qubit 15 is replaced by qubit 16, qubit 17 is replaced by qubit 1, qubit 18 is replaced by qubit 3, qubit 19 is replaced by qubit 4, qubit 20 is replaced by qubit 12, qubit 21 is replaced by qubit 16, qubit 22 is replaced by qubit 14, qubit 23 is replaced by qubit 14, qubit 24 is replaced by qubit 6.}
	\label{fig:SurfaceCodeRotated}
\end{figure}

As a final relevant explicit example, we consider a surface code that can correct two physical errors. While the surface code depicted in Fig.~\ref{fig:SurfaceCode} would contain 41 data qubits and 40 ancillary qubits in order to detect and cure two physical errors,  there are simple methods to reduce the number of qubits while keeping the distance the same. \cite{Horsman} Such modifications are important for small-scale implementations of surface codes in a near future; indeed there is clearly an intention to realize a quantum code that requires the smallest possible amount of resource.  Here we follow the approach of Ref.~\onlinecite{Horsman} and consider the rotated surface code of Fig.~\ref{fig:SurfaceCodeRotated}. The qubits that are part of the rotated code reside inside the black square and some of the boundary ancillary qubits are also incorporated to measure the boundary stabilizers. As was shown in Ref.~\onlinecite{Horsman}, such rotated surface code can correct two logical qubit errors although it possesses many fewer than 41 data qubits, in fact it consists only of 25 data qubits and 24 ancillary qubits. Furthermore, one can do slightly better by requiring not each stabilizer to have its individual ancillary qubit but instead by re-using an ancillary qubit to measure several stabilizer operators. We thus remove the 14 ancillary qubits with a (yellow) cross in Fig.~\ref{fig:SurfaceCodeRotated}.

\begin{figure}[h!]
	\centering
		\includegraphics[width=0.5\textwidth]{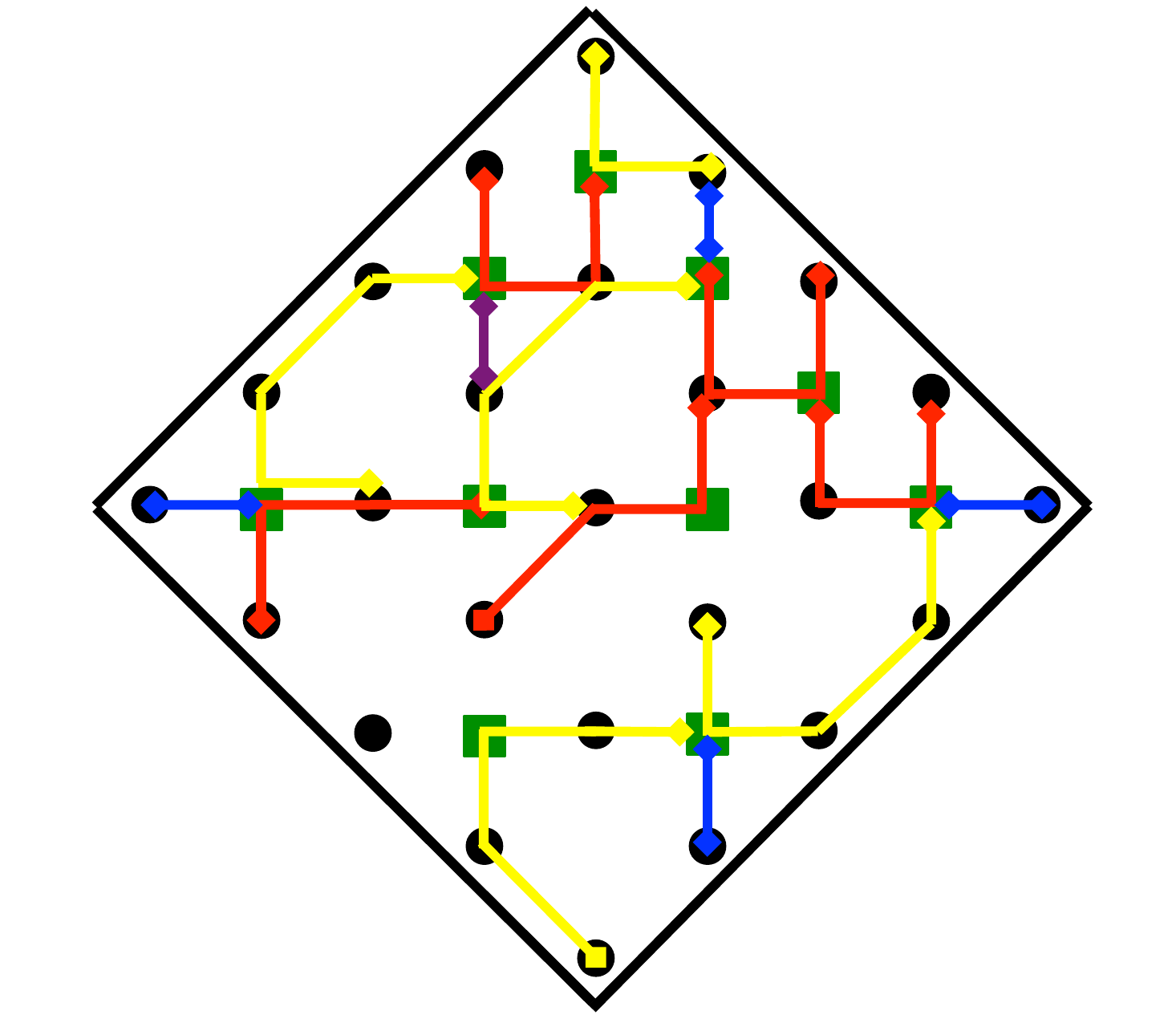}
	\caption{Optimal TLR scheme for the rotated surface code of Fig.~\ref{fig:SurfaceCodeRotated}, able to detect and correct two physical qubit errors. The TLRs are represented by solid lines with starting and ending points depicted by squares.}
	\label{fig:SurfaceCodeRotatedOptimal}
\end{figure}

We can now solve the linear binary program and find the optimal architecture given in Fig.~\ref{fig:SurfaceCodeRotatedOptimal}.

\section{Conclusions}
In this work we have developed a methodology to find optimal architectures for quantum codes. Our starting point is to consider a two-dimensional lattice of transmon qubits that interact with each others over moderate distances by coupling them to transmission line resonators. For each layout, we define a cost that allows to designate an optimal scheme. We show that finding such optimal scheme reduces to solve standard binary programs. What optimal means here depends obviously on the choice of a cost function. While we decided to choose to optimize over the total length of transmission line resonators for the Fibonacci and surface codes, our formalism is general enough to be straightforwardly applicable to many other codes and cost functions. In particular, we show how to apply our method to a restricted set of qubit and couplers that can be scaled up to a large two-dimensional structure.
\section{Acknowledgments}
We are happy to thank N. Breuckmann and B. Criger for useful discussions. We are grateful for support from the Alexander von Humboldt foundation and QALGO.

\end{document}